\documentclass[twoside,twocolumn,9pt]{article}
\usepackage{extsizes}
\usepackage[super,sort&compress,comma]{natbib} 
\usepackage[version=3]{mhchem}
\usepackage[left=1.5cm, right=1.5cm, top=1.785cm, bottom=2.0cm]{geometry}
\usepackage{balance}
\usepackage{widetext}
\usepackage{times,mathptmx}
\usepackage{sectsty}
\usepackage{graphicx} 
\usepackage{lastpage}
\usepackage[format=plain,justification=raggedright,singlelinecheck=false,font={stretch=1.125,small,sf},labelfont=bf,labelsep=space]{caption}
\usepackage{float}
\usepackage{fancyhdr}
\usepackage{fnpos}
\usepackage[english]{babel}
\usepackage{array}
\usepackage{droidsans}
\usepackage{charter}
\usepackage[T1]{fontenc}
\usepackage[usenames,dvipsnames]{xcolor}
\usepackage{setspace}
\usepackage[compact]{titlesec}

\usepackage{lpic}
\graphicspath{ {Pictures/} }

\definecolor{cream}{RGB}{222,217,201}

\begin{document}

\pagestyle{fancy}
\thispagestyle{plain}
\fancypagestyle{plain}{

%%%HEADER%%%
%\fancyhead[C]{\includegraphics[width=18.5cm]{head_foot/header_bar}}
%\fancyhead[L]{\hspace{0cm}\vspace{1.5cm}\includegraphics[height=30pt]{head_foot/journal_name}}
%\fancyhead[R]{\hspace{0cm}\vspace{1.7cm}\includegraphics[height=55pt]{head_foot/RSC_LOGO_CMYK}}
\renewcommand{\headrulewidth}{0pt}
}
%%%END OF HEADER%%%

%%%PAGE SETUP - Please do not change any commands within this section%%%
\makeFNbottom
\makeatletter
\renewcommand\LARGE{\@setfontsize\LARGE{15pt}{17}}
\renewcommand\Large{\@setfontsize\Large{12pt}{14}}
\renewcommand\large{\@setfontsize\large{10pt}{12}}
\renewcommand\footnotesize{\@setfontsize\footnotesize{7pt}{10}}
\makeatother

\renewcommand{\thefootnote}{\fnsymbol{footnote}}
\renewcommand\footnoterule{\vspace*{1pt}% 
\color{cream}\hrule width 3.5in height 0.4pt \color{black}\vspace*{5pt}} 
\setcounter{secnumdepth}{5}

\makeatletter 
\renewcommand\@biblabel[1]{#1}            
\renewcommand\@makefntext[1]% 
{\noindent\makebox[0pt][r]{\@thefnmark\,}#1}
\makeatother 
\renewcommand{\figurename}{\small{Fig.}~}
\sectionfont{\sffamily\Large}
\subsectionfont{\normalsize}
\subsubsectionfont{\bf}
\setstretch{1.125} %In particular, please do not alter this line.
\setlength{\skip\footins}{0.8cm}
\setlength{\footnotesep}{0.25cm}
\setlength{\jot}{10pt}
\titlespacing*{\section}{0pt}{4pt}{4pt}
\titlespacing*{\subsection}{0pt}{15pt}{1pt}
%%%END OF PAGE SETUP%%%

%%%FOOTER%%%
\fancyfoot{}
%\fancyfoot[LO,RE]{\vspace{-7.1pt}\includegraphics[height=9pt]{head_foot/LF}}
%\fancyfoot[CO]{\vspace{-7.1pt}\hspace{13.2cm}\includegraphics{head_foot/RF}}
%\fancyfoot[CE]{\vspace{-7.2pt}\hspace{-14.2cm}\includegraphics{head_foot/RF}}
\fancyfoot[RO]{\footnotesize{\sffamily{1--\pageref{LastPage} ~\textbar  \hspace{2pt}\thepage}}}
\fancyfoot[LE]{\footnotesize{\sffamily{\thepage~\textbar\hspace{3.45cm} 1--\pageref{LastPage}}}}
\fancyhead{}
\renewcommand{\headrulewidth}{0pt} 
\renewcommand{\footrulewidth}{0pt}
\setlength{\arrayrulewidth}{1pt}
\setlength{\columnsep}{6.5mm}
\setlength\bibsep{1pt}
%%%END OF FOOTER%%%

%%%FIGURE SETUP - please do not change any commands within this section%%%
\makeatletter 
\newlength{\figrulesep} 
\setlength{\figrulesep}{0.5\textfloatsep} 

\newcommand{\topfigrule}{\vspace*{-1pt}% 
\noindent{\color{cream}\rule[-\figrulesep]{\columnwidth}{1.5pt}} }

\newcommand{\botfigrule}{\vspace*{-2pt}% 
\noindent{\color{cream}\rule[\figrulesep]{\columnwidth}{1.5pt}} }

\newcommand{\dblfigrule}{\vspace*{-1pt}% 
\noindent{\color{cream}\rule[-\figrulesep]{\textwidth}{1.5pt}} }

\makeatother
%%%END OF FIGURE SETUP%%%

%%%TITLE, AUTHORS AND ABSTRACT%%%
\twocolumn[
  \begin{@twocolumnfalse}
\vspace{3cm}
\sffamily
%\begin{tabular}{m{4.5cm} p{13.5cm} }

%\includegraphics{head_foot/DOI} & \noindent\LARGE{\textbf{Freezing and melting line invariants of the Lennard-Jones system$^\dag$}} \\%Article title goes here instead of the text "This is the title"
\noindent\LARGE{\textbf{Freezing and melting line invariants of the Lennard-Jones system$^\dag$}} 
\vspace{1cm}	%added <-----------------------------------------
%\vspace{0.3cm} & \vspace{0.3cm} \\

\noindent\large{Lorenzo Costigliola,$^{\ast}$ Thomas B. Schr{\o}der, and Jeppe C. Dyre} \\%Author names go here instead of "Full name", etc.
\vspace{1.5cm}	%added <-----------------------------------------

\noindent\normalsize{The invariance of several structural and dynamical properties of the Lennard-Jones (LJ) system along the freezing and melting lines is interpreted in terms of the isomorph theory.
First the freezing/melting lines for LJ system are shown to be approximated by isomorphs. 
Then we show that the invariants observed along the freezing and melting isomorphs  
are also observed on other isomorphs in the liquid and crystalline phase. 
Structure is probed by the radial distribution function and the structure factor and dynamics is probed by the mean-square displacement, the intermediate scattering function, and the shear viscosity. Studying these properties by reference to the isomorph theory explains why known single-phase melting criteria holds, e.g., the Hansen-Verlet and the Lindemann criterion, and why the Andrade equation for the viscosity at freezing applies, e.g., for most liquid metals. Our conclusion is that these empirical rules and invariants can all be understood from the isomorph theory 
and that the invariants are not peculiar to the freezing and melting lines, but hold along all isomorphs.} \\%The abstrast goes here instead of the text "The abstract should be..."

%\end{tabular}

 \end{@twocolumnfalse} \vspace{0.6cm}

  ]
%%%END OF TITLE, AUTHORS AND ABSTRACT%%%

%%%FONT SETUP - please do not change any commands within this section
\renewcommand*\rmdefault{bch}\normalfont\upshape
\rmfamily
\section*{}
\vspace{-1cm}

%%%FOOTNOTES%%%

\footnotetext{\textit{~``Glass and Time'', IMFUFA, Department of Sciences, Roskilde University, Postbox 260, DK-4000 Roskilde, Denmark. E-mail: lorenzo.costigliola@gmail.com}}

\twocolumn

\section{Introduction}

In this work several freezing line and melting line invariants, both structural and dynamical,
of the Lennard-Jones (LJ) system \cite{LJ1924} are derived from the isomorph theory \cite{paper4} and validated in computer simulations.
The existence of invariances along isomorphs is used to explain the Hansen-Verlet and Lindemann freezing/melting criteria as well as the Andrade equation for the freezing viscosity for the LJ system. 

The phase transition from liquid to crystal and {\it vice versa} is not yet completely understood \cite{Cahn1986,Oxtoby1990,Cahn2001}. Reasons for searching for a better understanding of freezing/melting invariants are many.
One is the possibility of using freezing/melting invariance to evaluate specific system properties under conditions not easily accessible by experiments.
An example could be the estimation of liquid iron's viscosity at Earth-core pressure and temperature conditions, a quantity that is necessary for developing reliable geophysical models for the core\cite{Poirier1988,Poirier2000,Stacey2010}.

Many theories have been proposed to explain freezing and melting\cite{Ubbelohde1965,Wallace2002} and why certain quantities are often invariant along the freezing and melting lines.
Examples of such invariants are the excess entropy, the constant-volume entropy difference between liquid and solid on melting\cite{Tallon1989,Tallon1980,Stishov1975}, the height of the first peak of the static structure factor on freezing (the Hansen-Verlet freezing criterion\cite{Hansen1969,Hansen1970}), and the viscosity of liquid metals on freezing when made properly dimensionless\cite{Andrade1931,Andrade1934a,Andrade1934b}.
The Lindemann\cite{Lindemann1910,Gilvarry1956} melting criterion states that a crystal melts when
the mean vibrational displacement of atoms from their lattice position exceeds $0.1$ of the mean inter-atomic distance, 
independent of the pressure.
This is equivalent to invariance of $<u^2>/r^2_m$ along the melting line\cite{Gilvarry1956}, 
where $<u^2>$ is the atomic root-mean-squared vibrational amplitude
and $r_m$ is the nearest neighbor distance.
The most common approaches for explaining such invariants attempt to connect them to the kinetics of the freezing/melting process.
For instance, going back to Born it has been suggested that a crystal becomes mechanically unstable when $<u^2>/r^2_m$ exceeds a certain number\cite{Ubbelohde1965}. 
From this perspective, it is not easy to understand why these invariants do not hold for all systems.
It is also difficult to understand why related invariants hold on specific curves in the liquid state.
Thus, in an extension of what happens along the melting line of, e.g., the Lennard-Jones system, the radial distribution function is invariant along the curves at which the excess entropy $S_{ex}$ is equal to the two-body entropy $S_2$\cite{Saija2001}.
Diffusivity is also constant, in appropriate units, along constant $S_{ex}$ curves\cite{Dzugutov1996}, implying (from the Stokes-Einstein relation) invariance of the viscosity in appropriate units along these curves.
This relation between viscosity and excess entropy was recently confirmed by high-pressure measurements\cite{Abramson2014}.

A possible explanation of the invariants along the freezing and melting lines, as well as along other well-defined curves in the thermodynamic phase diagram, is given by the isomorph theory\cite{paper1,paper4,hrhoTrond,IsomorphCrystal}.
According to it\cite{simpleliquid} a large class of liquids exists for which structure and dynamics are invariant to a good approximation along the constant-excess-entropy curves. 
These curves are termed isomorphs, and the liquids which conform to the isomorph theory are now called Roskilde-simple (R) liquids\cite{simpleliquid,Arno2015,Isomorph2.0,Bailey2014,quasiuniNVU,quasiuni2} (the original name ``strongly correlating'' caused confusion due to the existence of strongly correlated quantum systems). Liquids belonging to this class are easily identified in computer simulations because they exhibit strong correlations between their thermal-equilibrium fluctuations of virial and potential energy in the $NVT$ ensemble\cite{paper1,paper2}.
The isomorph theory offers not only the possibility of explaining the freezing/melting invariants without reference to the actual mechanisms of the freezing/melting process itself; by evaluating the virial potential-energy correlation coefficient it also provides a way to predict whether these invariants hold for a given liquid.

The main features of the isomorph theory are summarized in Sec. \ref{sec:isomorph} where it is also shown how to identify the isomorphs of the LJ system. 
This is followed by a short section describing technical details of the simulations performed. 
The isomorph equations are used in Sec. \ref{sec:melting} to show that the freezing line can be approximated by an isomorph, termed the freezing isomorph, without need of any fitting.
Sec. \ref{sec:resultsfreezing} deals with freezing invariants, the Hansen-Verlet criterion\cite{Hansen1969,Hansen1970}, 
and Andrade's freezing viscosity equation\cite{Andrade1931,Andrade1934a,Andrade1934b};
Sec. \ref{sec:resultsmelting} focuses on melting line invariants of FCC LJ crystal and their connection with 
the Lindemann criterion\cite{Lindemann1910}. 
The last section discusses the differences between isomorph theory and other approaches used to describe liquid invariances 
in the past years and summarizes the main results of this work.

\section{Isomorphs} \label{sec:isomorph}
An R system is characterized by strong correlations between virial and potential energy equilibrium fluctuations 
in the NVT ensemble\cite{paper1,paper2}, i.e., 
by a virial potential-energy equilibrium correlation coefficient $R(\rho, T)$ greater than $0.9$:
\begin{equation}\label{eq:Req}
\begin{aligned}
R(\rho, T)=\frac{\langle \Delta W \Delta U\rangle}{\sqrt{\langle (\Delta W)^2 \rangle \langle (\Delta U)^2 \rangle}}>0.9 \quad \text{.}
\end{aligned}
\end{equation}
Here $\Delta$ denotes the instantaneous deviations from the equilibrium mean value and the brackets denote $NVT$ ensemble averages, $W$ the virial, $U$ the internal energy and $(\rho, T)$ the density and temperature of the system.
When such strong correlations are present, the theory predicts the existence of curves in the thermodynamic phase diagram along which several structural, dynamical, and thermodynamical properties are invariant 
\cite{paper1, paper2, paper3, paper4, paper5} when expressed in reduced units; these curves are termed isomorphs\cite{paper4}.

Reduced quantities (marked by a tilde) are defined as follows.
Distances are measured in units of $\rho^{-1/3}$, energies in units of $k_B T$, and time in units of $m^{1/2}(k_B T)^{-1/2}\rho^{-1/3}$, 
where $m$ is the average particle mass (for Brownian dynamic a different time unit applies\cite{paper4}). 
These reduced units should not be confused with the so-called Lennard-Jones (LJ) units.
We use the latter units below for reporting quantities like the temperature and density.

By definition an isomorph has the following property: 
for any two configurations ${\bf{R}}_{1}\equiv ({\bf{r}}_1^{(1)}, ..., {\bf{r}}_N^{(1)})$ 
and ${\bf{R}}_{2}\equiv ({\bf{r}}_1^{(2)}, ..., {\bf{r}}_N^{(2)})$
\begin{equation}
\label{eq:Bfactors}
\begin{aligned}
\quad \rho_1^{1/3} {\bf{R}}_{1}=\rho_2^{1/3} {\bf{R}}_{2} \quad \Rightarrow \quad P({\bf{R}}_{1})=P({\bf{R}}_{2}) 
\end{aligned}
\end{equation} 
where ${\bf{r}}_i$ is the position vector of the particle $i$, $N$ is the number of particles and 
$P({\bf{R}}_{i})$ is the Boltzmann statistical weight of configuration ${\bf{R}}_{i}$ at 
the relevant thermodynamic state point on the isomorph\cite{paper1}.
In other words, configurations that are identical in reduced units ($\tilde{\bf{R}}\equiv\rho^{1/3} {\bf{R}}$)
have proportional Boltzmann factors.

The isomorph theory is exact only for systems with an Euler-homogeneous potential energy function, 
for instance inverse-power-law (IPL) pair-potential systems\cite{paper1, paper2}. 
However, the theory can be used as a good approximation for the wide class of systems. 
Examples of models that are R liquids\cite{simpleliquid} in part of their thermodynamic phase diagram, 
in liquid and solid state\cite{IsomorphCrystal}, 
are the standard and generalized Lennard-Jones systems (single-component as well as multi-component) 
\cite{paper4, paper5, hrhoLasse}, 
systems interacting via the exponential pair potential \cite{Bacher2014}, 
systems interacting via the Yukawa potential \cite{ArnoPhD,Arno2015}.
R systems include also some molecular systems like, e.g., the asymmetric dumbbell models\cite{Trond2012},  the Lewis-Wahnstr{\"{o}}m's three-site model of OTP \cite{Trond2012}, the seven-site united-atom model of toluene\cite{paper1}, the EMT model of liquid Cu \cite{paper1} and the rigid-bond Lennard-Jones chain model \cite{Arno2014}.
Predictions of the isomorph theory have been shown to hold in experiments on glass-forming van der Waals liquids by Gundermann \emph{et al}\cite{Gundermann2011}, by Roed \emph{et al}\cite{Roed2013}, and by Xiao \emph{et al}\cite{Xiao2015}.
Power-law density scaling\cite{Roland2005}, which is often observed in experiments on viscous liquids, 
can be explained by the isomorph theory\cite{hrhoLasse}.

Isomorphic scaling, i.e., the invariance along isomorphs of many reduced quantities deriving from the identical statistical weight of scaled configurations\cite{paper4}, does not hold for all reduced quantities. 
For example, the reduced-unit free energy and pressure are not invariant, whereas the excess entropy, reduced structure, 
and reduced dynamics are all isomorph invariant\cite{paper4}.
These invariances follow from the invariance along isomorphs of Newtonian and Brownian equations of motion 
in reduced units for R liquids\cite{paper4}.

For an R the system at a given reference state point $(\rho_0, T_0)$, it is possible to build an isomorph 
starting from that point\cite{paper4}.
For R systems, a function $h(\rho)$ exists which relates the state point $(\rho_0, T_0)$ 
to any other state point $(\rho, T)$ along the same isomorph\cite{hrhoTrond, hrhoLasse} by the identity:
\begin{equation} \label{eq:T(rho)1}
\begin{aligned}
\frac{h(\rho)}{T}=\frac{h(\rho_0)}{T_0} \quad \text{.}
\end{aligned}
\end{equation}
The functional form of $h(\rho)$ depends on the interaction potential, 
and only for simple systems is it possible to find an analytical expression. 
As shown by Ingebrigtsen \emph{et al}\cite{hrhoTrond} and B{\o}hling \emph{et al}\cite{hrhoLasse}, 
if the pair potential is a sum of inverse-power laws involving the exponents $n_i$ ($i=1,...,N$), 
$h(\rho)$ can be expressed in the following way:
\begin{equation} \label{eq:h(rho)1}
h(\rho)=\sum^N_{i=1} \alpha_i \left(\frac{\rho}{\rho_0}\right)^{{n_i}/3}  \quad \text{.}
\end{equation}
For a LJ system the pair potential is the well-known
\begin{equation}\label{eq:LJpotential}
v(r)=4\varepsilon\left((r/\sigma)^{-12}-(r/\sigma)^{-6}\right)\,
\end{equation}
so only two IPL exponents, $12$ and $6$, are involved.
It is not difficult to show that\cite{hrhoTrond, hrhoLasse} for the LJ system, $h(\rho)$ is given by
\begin{equation} \label{eq:h(rho)2}
h(\rho)=\left(\frac{\gamma_0}{2}-1\right)\left(\frac{\rho}{\rho_0}\right)^4-\left(\frac{\gamma_0}{2}-2\right)\left(\frac{\rho}{\rho_0}\right)^2 \\
\end{equation}
where $\gamma_0$ is the so-called density-scaling exponent at the reference state point defined by the canonical averages
\begin{equation} \label{eq:gamma}
\gamma_0(\rho_0, T_0)={\frac{\langle\Delta W \Delta U\rangle}{\langle (\Delta U)^2\rangle} \vline}_{(\rho_0,T_0)}  \quad \text{.}
\end{equation}
Equation (\ref{eq:h(rho)2}) is easily derived from applying $\gamma =d\ln h/d\ln\rho$\cite{hrhoTrond} at the reference state point to Eq. (\ref{eq:h(rho)1}),
adopting the normalization $h(\rho)=1$. 
The correlation coefficient $R$ of the LJ system increases with increasing temperature 
and with increasing density \cite{paper1}; 
this means that if the LJ system is an R liquid at the reference state point $(\rho_0,T_0)$, 
it will be strongly correlating also at higher densities on the isomorph through $(\rho_0,T_0)$.

Recently isomorph theory have been reformulated starting from the assumption that for any couple of configurations of a R systems, the potential energies obeys the relation
\begin{equation}
U({\bf{R}}_{1}) < U({\bf{R}}_{2}) \implies U(\lambda{\bf{R}}_{1}) < U(\lambda{\bf{R}}_{2} )
\end{equation}
when the configurations are scaled to a different density\cite{Isomorph2.0}. All the results described in this section can be derived from this simple scaling rule.

\section{Simulation details} \label{sec:simulations}

This work presents results of molecular dynamics simulations of single-component LJ system performed using the GPU code RUMD\cite{rumdpaper}. 
For each liquid state point an NVT simulation was used to obtain structure and dynamics,
while a SLLOD simulation\cite{EvansMorriss2008, DaivisTodd2006, EvansHeyes1990} was used to find the viscosity.  
The simulations were carried out using a shifted-potential cutoff at $2.5 \sigma$.
In the simulations the LJ parameters were set to unity, i.e., $\sigma=1.0$ and $\epsilon=1.0$.
The time step was adjusted with increasing temperature along an isomorph to keep the reduced time step constant, equal to $0.001$ for all simulations. 
For instance, the time step is $0.001$ in LJ units for a simulation at $\rho=1.0$ and $T=1.0$.
At every state point the system was simulated for $5\cdot 10^8$ timesteps, which takes about $20$ hours (in the case of SLLOD simulations) on a modern GPU card (Nvidia GTX 780 Ti).
The NVT simulations used to calculate $\gamma$ and $R$ at the starting state point for any isomorph 
ran for $10^{10}$ time steps in order to get good statistics for $\gamma$.
In the NVT simulations of the FCC LJ crystal, the thermostat time constant was kept constant in reduced units.
The value for the reduced thermostat constant is $0.4$.
Details on how to obtain viscosity from SLLOD simulations can be found in the Appendix.
In the liquid phase and along the freezing line, 1000 LJ particles were simulated;
for the FCC LJ crystal 4000 LJ particles were simulated.

\section{The Freezing line}  \label{sec:melting}
As mentioned in section \ref{sec:isomorph}, along an isomorph scaled configurations have the same statistical weight.
This implies that the freezing and melting lines of an R liquid are isomorphs: 
consider a state point of the fluid state in which, therefore, the disordered configurations are the most likely,
and another state point in which the system is in a crystalline phase.
Since in the latter case the ordered configurations are most likely, these two state points cannot be on the same isomorph.
It follows that the freezing and melting lines cannot be crossed by an isomorph (in the region where the system is a R system), i.e., 
in both the liquid and crystalline regions isomorphs must be parallel to the freezing and melting lines, respectively.
In particular, these lines are themselves isomorphs.
This statement follows from assuming that the physically relevant states obey the isomorph scaling condition\cite{paper4}.
\begin{figure}
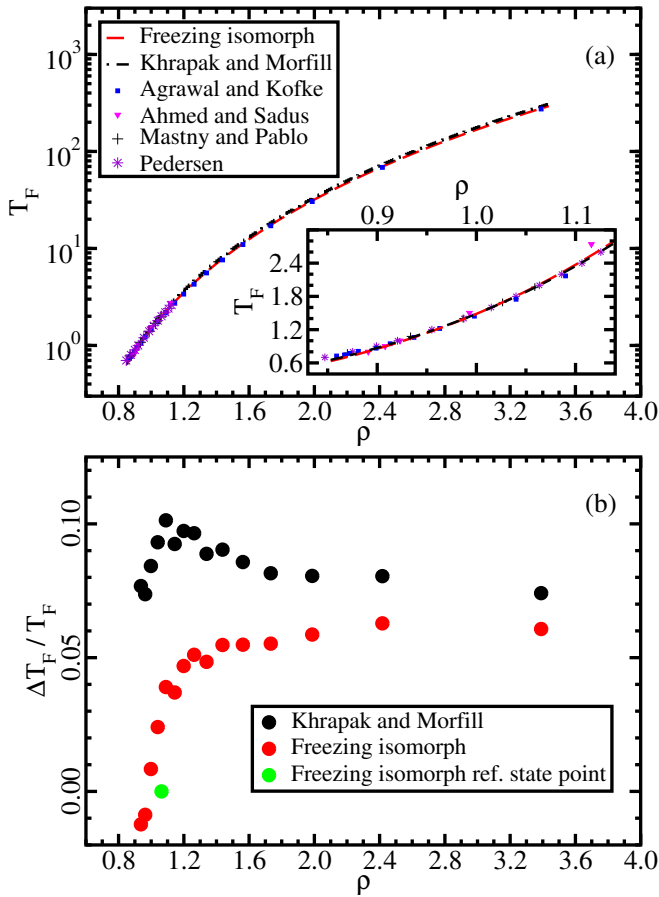

\centering
  \resizebox{9cm}{!}{
  \begin{tabular}{l}
  \begin{lpic}[t(0.5mm)]{MeltingNew(0.5)}\end{lpic} \\ %0.29
  \begin{lpic}[]{3_meltingLines(0.5)}\end{lpic}
  \end{tabular}}%}
  \captionof{figure}{
  Freezing line of the LJ system. In (a) the isomorph approximation to the freezing line is marked by the red line and 
  the Khrapak and Morfill approximation\cite{Khrapak2011} by the black line; 
  freezing state points obtained in the past years using various techniques 
  are shown by symbols\cite{Ahmed2013,Mastny2007,ulfpinning,Agrawal1995}.
  Both approximations reproduce the data points well; the inset focuses on low densities.
  In (b) the relative difference between Agrawal and Kofke freezing-temperature data\cite{Agrawal1995} and the two approximations
  is shown. The isomorph approximation gives smaller deviations from the simulation data.
  The main advantage of approximating the freezing line by an isomorph lies, however, in the possibility of predicting 
  the full freezing line from knowledge of a single freezing state point. 
  }
  \label{fig:iso}
\end{figure}

The LJ system is an R liquid so its freezing line is approximately an isomorph.
This was first confirmed by Schr{\o}der \emph{et al} \cite{paper5} using data from computer simulations by Ahmed and Sadus \cite{Ahmed2013} 
and Mastny and de Pablo \cite{Mastny2007}, and subsequently by Pedersen \cite{ulfpinning} with data obtained by his 
interface-pinning method\cite{pinningmethod}. 
Recently, the approximate isomorph nature of the freezing line has been documented in detail by Heyes \emph{et al}\cite{Heyes2015a,Heyes2015b}. 
The quoted papers all focus on densities fairly close to unity (in LJ units). 
From the fact that the freezing line is an isomorph it is possible to understand the invariance along the freezing line of several properties, as recently was shown by Heyes \emph{et al}\cite{Heyes2015a}, who studied the invariance of the reduced-unit radial distribution function, 
mean force, Einstein frequency, self-diffusion coefficient, and linear viscoelasticity of an LJ liquid along
the freezing line, for densities around unity. 
All these quantities were found to be approximately invariant, as predicted by the isomorph theory.

In this section the validity of an equation for the freezing line of the LJ system obtained from isomorph theory 
is checked over a considerably wider range of temperatures and densities than previously studied.
In section \ref{sec:resultsfreezing}, the results of Heyes \emph{et al}\cite{Heyes2015a} on structural and dynamic invariants
are extended to a wide range of densities along the freezing line.

In Fig. \ref{fig:iso} the agreement between the freezing isomorph and the freezing line is shown to 
hold in the whole range of temperatures and densities studied by Agrawal and Kofke \cite{Agrawal1995}.
The red line in figure \ref{fig:iso} is the prediction from the isomorph theory; 
this line is built by starting from  the freezing point $T_0=2.0$ and $\rho_0=1.063$, 
obtained by Pedersen\cite{ulfpinning}.
The correlation coefficient $R$ and the scaling parameter $\gamma$ at the state point $(\rho_0,T_0)$ are: 
\begin{equation}\label{eq:Rgammavalues}
\begin{aligned}
R_0=0.995 \quad \text{,} \quad \gamma_0=4.907 \quad \text{.}
\end{aligned}
\end{equation}

Using Eqs. (\ref{eq:T(rho)1}) and (\ref{eq:h(rho)2}) and this value for $\gamma_0$, 
it is possible to build the freezing isomorph from
\begin{equation} \label{eq:T(rho)2}
\begin{aligned}
T_F(\rho) = A_{F} \rho^4 - B_{F} \rho^2
\end{aligned}
\end{equation}
where $T_F$ is the freezing temperature and 
$A_{F}=2.27$, $B_{F}=0.80$ is found from the reference state-point information given Eqs. (\ref{eq:h(rho)2}) 
and (\ref{eq:Rgammavalues}).
The same power-law dependence for the LJ freezing line was obtained in 2009 by Khrapak and Morfill\cite{Khrapak2011} and, 
in fact, long ago by Rosenfeld from his ``additivity of melting temperatures'' 
(derived by reference to the hard-sphere system)\cite{Rosenfeld1976,Rosenfeld1976b}.
This is consistent with isomorph theory because Rosenfeld's rule can be derived from the quasi-universality 
of single-component R liquids\cite{quasiuniNVU, quasiuni2}. 

Fitting to the same simulation for the freezing line as referenced above\cite{Ahmed2013, Mastny2007, Agrawal1995}, 
Khrapak and Morfill \cite{Khrapak2011} found the following values for the coefficients: $A=2.29$ and $B=0.71$.
The line obtained inserting these values of $A$ and $B$ into Eq. (\ref{eq:T(rho)2}) is shown in Fig. \ref{fig:iso} (a) (black dots).
There is a significant difference in the second coefficient between the two equations. 
The second coefficient of Khrapak and Morfill is obtained using data for the triple point 
which may explain the difference; in that region the isomorph theory does not provide a good approximation 
for the freezing line of the LJ system, as Pedersen recently showed \cite{ulfpinning}. 
Nevertheless, the two curves are close to each other.
The freezing isomorph provides a slightly better prediction of freezing temperatures at any density when compared to the Khrapak and Morfill fit 
(inset in Fig. \ref{fig:iso} (a) and Fig. \ref{fig:iso} (b)).

The main result of this section is that isomorph theory provides a technique for 
approximating the freezing line of an R liquid from simulations at a single state point, i.e., without any fitting, 
and that this approximation is valid over a wide range of densities. 
The relative difference between the predicted freezing temperature and the one obtained from computer simulations
\cite{Agrawal1995} is about $6\%$ for density change of more than a factor $3$ and 
temperature change of more that a factor $100$, as shown in Fig. \ref{fig:iso}.
The isomorph theory allows therefore to estimate the freezing temperatures with small relative uncertainties, 
and it may be useful for estimating the freezing temperatures in the high density regimes, 
where is difficult to perform direct experiments, for real liquids which are R liquids in the relevant part of the phase diagram.

\section{Invariants along the freezing line} \label{sec:resultsfreezing}

In this section we discuss different invariants along the freezing isomorph 
as well as another isomorph "parallel" to it in the liquid state, generated from the state point $(\rho, T)=(1.063, 4.0)$.
It is demonstrated that invariants originally proposed for the freezing line are found also along the liquid isomorph.
Along the two isomorphs investigated, the excess pressure in reduced units is also evaluated (Fig. \ref{fig:redPandVisco}(a)).
This quantity is invariant for any IPL system, but not for the LJ system.
In the framework of isomorph theory, it is well understood why some quantities are invariant, e.g., the reduced viscosity, 
while others are not, e.g., the reduced pressure\cite{paper4}.
This shows that the scaling properties studied in this work are not simply the consequences of an effective IPL scaling.
Note also that it is necessary to go to quite high densities before $\gamma\approx4$, 
as shown in Fig. \ref{fig:redPandVisco}(b). 
In the same figure, the correlation coefficient $R$ and the reduced viscosity are plotted as a function of density 
along the freezing isomorph. The reduced viscosity is predicted to be invariant\cite{paper4}. 
For $\rho>1.1$ the reduced viscosity is invariant to a good approximation.
At lower densities, the correlation coefficient $R$ decreases and the reduced viscosity begins to vary.

\subsection{Structure and the Hansen-Verlet freezing criterion}

\begin{figure*}
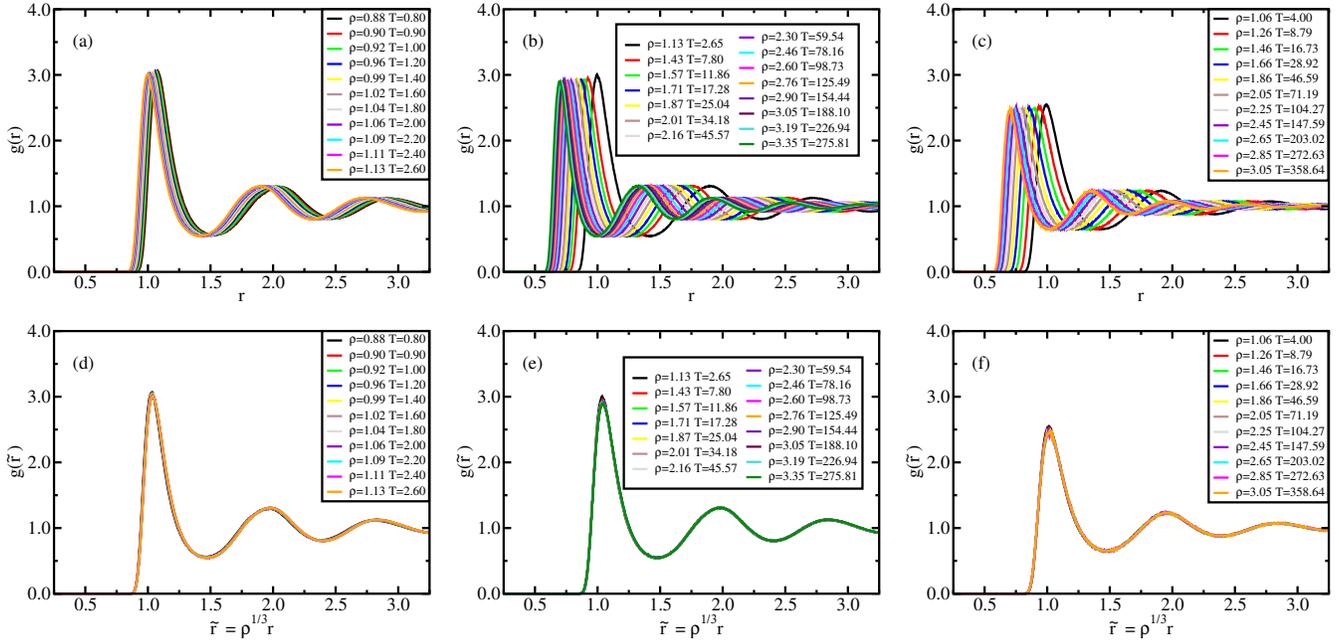

\centering
  \resizebox{18cm}{!}{
    \begin{tabular}{l l l}
      \begin{lpic}[]{ulf_melting_rdf(0.30)}\end{lpic} &
      \begin{lpic}[]{meltingIso_rdf(0.30)}\end{lpic} &
      \begin{lpic}[]{T4p0_rdf(0.30)}\end{lpic} \\
      \begin{lpic}[]{ulf_melting_rdf_reduced(0.30)}\end{lpic} &
      \begin{lpic}[]{meltingIso_rdf_reduced(0.30)}\end{lpic} &
      \begin{lpic}[]{T4p0_rdf_reduced(0.30)}\end{lpic} \\
    \end{tabular}}
  \caption{Liquid results. Radial distribution function along the Pedersen freezing line (a,d)\cite{ulfpinning}, 
  along the approximating freezing isomorph (b,e) and along an isomorph well within the liquid state (c,f); 
  in (a), (b), and (c), the RDFs are plotted as a function of distance in Lennard-Jones units,
  in (d), (e), and (f), the RDFs are plotted as a function of the reduced distance.
  It is worth noting that while in (a) and (d) the density change is only a few percent, 
  in the other figures density is changed of about a factor $3$. The same holds for Figs. \ref{fig:sq} - \ref{fig:ISF}.}
  \label{fig:RDF}
\end{figure*}

Figure \ref{fig:RDF} shows the radial distribution functions (RDF) $g(r)$ at different state points 
along the freezing line (a, d), the approximate freezing isomorph (b, e), and the liquid isomorph (c, f).
In Figs. \ref{fig:RDF} (a), (b), and (c), $g(r)$ is expressed as a function of the pair distance,
while in Figs. \ref{fig:RDF} (e), (f), and (g), the $g(r)$ is expressed as a function of the reduced distance, $\tilde{r}=\rho^{1/3} r$. 
When the RDFs are plotted in reduced units, they collapse onto master curves, as predicted by the isomorph theory.
The results for the freezing line confirm the recent findings of Heyes \emph{et al}\cite{Heyes2015}, 
who showed the same collapse albeit for a smaller density range.

\begin{figure*}
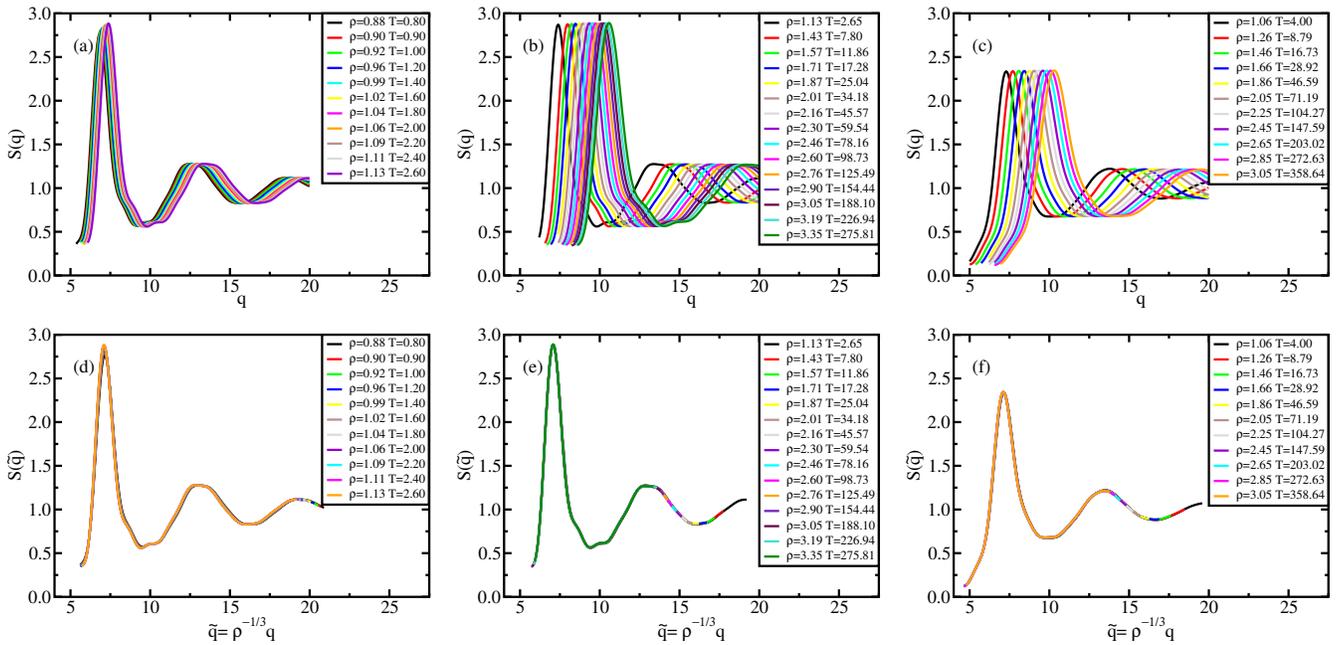

\centering
  \resizebox{18cm}{!}{
    \begin{tabular}{l l l}
      \begin{lpic}[]{ulf_melting_sq(0.29)}\end{lpic} & 
      \begin{lpic}[]{meltingIso_sq(0.29)}\end{lpic} &
      \begin{lpic}[]{T4p0_sq(0.29)}\end{lpic} \\
      \begin{lpic}[]{ulf_melting_sq_reduced(0.29)}\end{lpic} &
      \begin{lpic}[]{meltingIso_sq_reduced(0.29)}\end{lpic} &
      \begin{lpic}[]{T4p0_sq_reduced(0.29)}\end{lpic} \\
    \end{tabular}}
  \caption{Liquid results. Structure factor the along the Pedersen freezing line (a,d) \cite{ulfpinning},  
  along the approximate freezing isomorph (b,e), and along an isomorph well within the liquid state (c,f);
  in (a), (b), and (c), $S(q)$ is plotted as a function of wave vector in Lennard-Jones units, 
  in (d), (e), and (f), $S(q)$ is plotted as a function of reduced wave vector.}
  \label{fig:sq}
\end{figure*}

Starting from the invariance of $g(r)$ it is easy to show that the structure factor $S(q)$ is invariant
when considered as a function of the reduced wave vector,
\begin{eqnarray}\label{eq:sq}
\begin{aligned}
S(q)-1 &=& \rho \int_{V} {d \textbf{r} \quad e^{-i \textbf{q} \cdot \textbf{r}} g(r)} = \quad \quad \quad \quad \quad \quad\\ 
&=& \int_{\tilde{V}} {d \tilde{\textbf{r}} \quad e^{-i \left( \rho^{-1/3} \textbf{q} \right) \cdot \tilde{\textbf{r}}} g(\tilde{r})} = S(\tilde{q})-1 \quad \text{.} 
\end{aligned}
\end{eqnarray}
Structure factors $S(q)$ along the freezing line, the approximate freezing isomorph, 
and the liquid isomorph are shown in Fig. \ref{fig:sq}.
The invariance of the structure factor implies the Hansen-Verlet freezing criterion\cite{Hansen1969, Hansen1970} 
stating that the LJ system freezes when the height of the first peak of the structure factor reaches 
a definite value close to $3$ (equal to $2.85$ in the original work\cite{Hansen1969, Hansen1970}): 
if $S(q)$ is invariant along an isomorph, points which are on the same isomorph have the same height of the first peak. 
And since the freezing line for R liquids is well approximated by an isomorph, 
the invariance of $S(q)$ implies the validity of Hansen-Verlet freezing criterion. Figure \ref{fig:sq} confirms this.

\subsection{Dynamic invariants: mean-squared displacement, intermediate scattering function}

\begin{figure*}
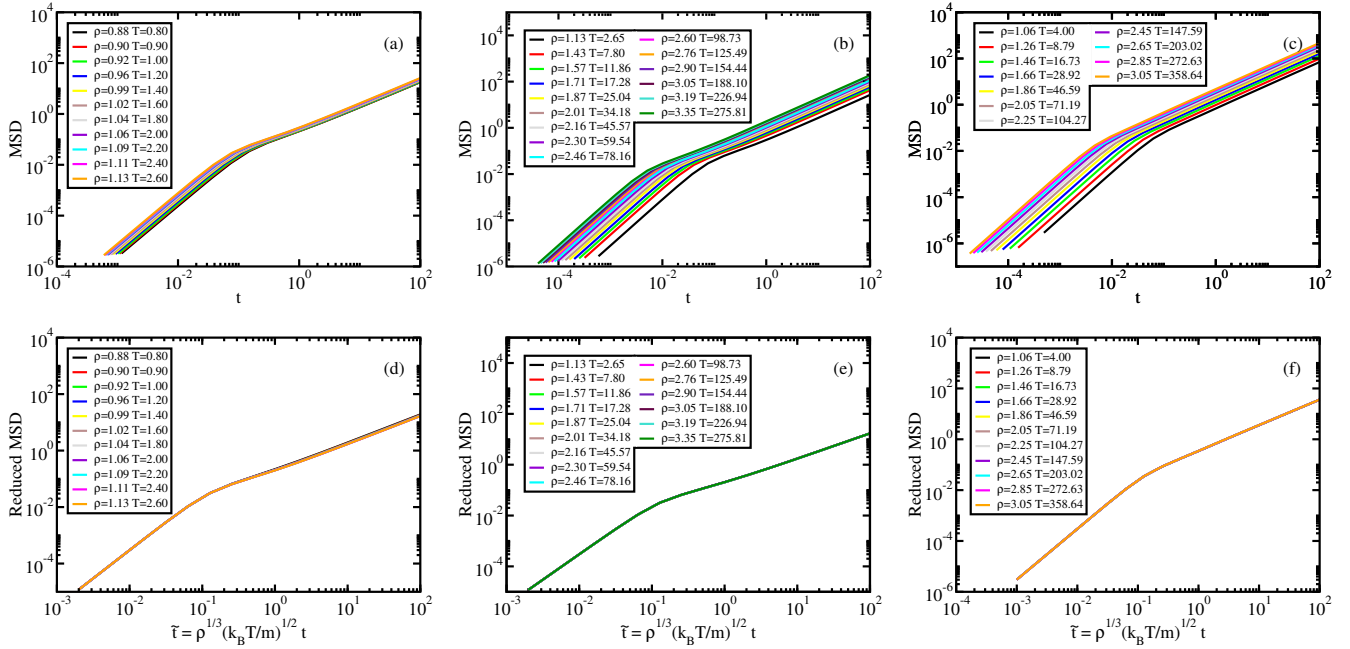

\centering
  \resizebox{18cm}{!}{
    \begin{tabular}{l l l}
  \begin{lpic}[]{ulf_melting_msd(0.29)}\end{lpic} &
  \begin{lpic}[]{meltingIso_msd(0.29)}\end{lpic} &
  \begin{lpic}[]{T4p0_msd(0.29)}\end{lpic} \\
  \begin{lpic}[]{ulf_melting_msd_reduced(0.29)}\end{lpic} &
  \begin{lpic}[]{meltingIso_msd_reduced(0.29)}\end{lpic} &
  \begin{lpic}[]{T4p0_msd_reduced(0.29)}\end{lpic} \\
    \end{tabular}}
  \caption{Liquid results. Mean-squared displacement along the Pedersen freezing line (a,d)\cite{ulfpinning}, 
  along the approximating freezing isomorph (b,e) and along another isomorph in the liquid state (c,f); 
  in (a), (b), and (c), the MSDs are plotted as a function of time in LJ units,
  in (d), (e), and (f), the reduced MSDs are plotted as a function of reduced time.}
  \label{fig:MSD}
\end{figure*}

\begin{figure*}
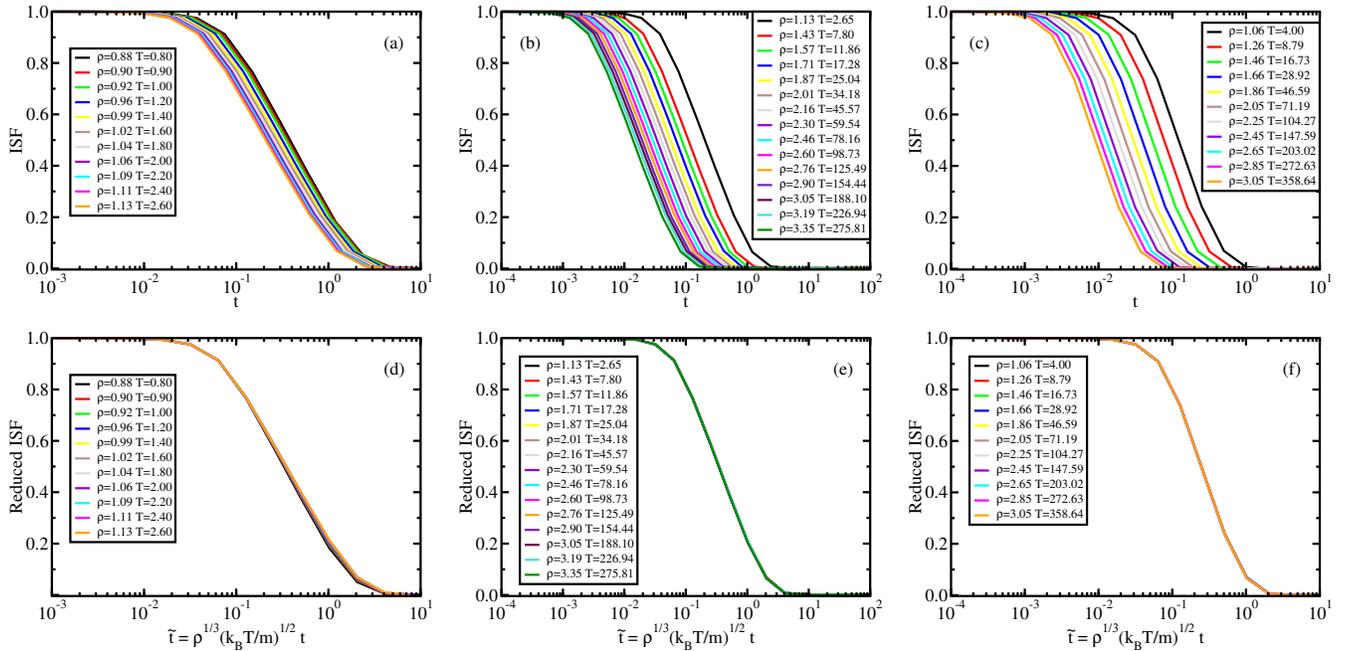

\centering
  \resizebox{18cm}{!}{
    \begin{tabular}{l l l}
  \begin{lpic}[]{ulf_melting_Fs(0.29)}\end{lpic} &
  \begin{lpic}[]{meltingIso_Fs(0.29)}\end{lpic} &
  \begin{lpic}[]{T4p0_Fs(0.29)}\end{lpic} \\
  \begin{lpic}[]{ulf_melting_Fs_reduced(0.29)}\end{lpic} &
  \begin{lpic}[]{meltingIso_Fs_reduced(0.29)}\end{lpic} &
  \begin{lpic}[]{T4p0_Fs_reduced(0.29)}\end{lpic} \\
    \end{tabular}}
  \caption{Liquid results. Self-intermediate scattering function along the Pedersen freezing line (a,d)\cite{ulfpinning}, 
  along the approximating freezing isomorph (b,e), and along another isomorph in the liquid state (c,f); 
  in (a), (b), and (c), the ISFs are plotted as a function of time in Lennard-Jones units,
  in (d), (e), and (f), the ISFs are plotted as a function of reduced time.
  All the ISFs correspond to the $q$ value of the first peak of $S(q)$, $q_{max}$.
  The quantity $\tilde{q}_{max}$ is invariant along an isomorph due to the invariance of $\tilde{S}(\tilde{q})$, 
  Eq. (\ref{eq:sq}).}
  \label{fig:ISF}
\end{figure*}

The dynamical behavior of the system is described by the mean-squared displacement (MSD)
and the self-intermediate scattering function (ISF). 
In Figs. \ref{fig:MSD} and \ref{fig:ISF}, the MSDs and ISFs are shown, respectively, as functions of non-reduced and reduced quantities. 
As for the structure, the curves collapse onto master curves.
\begin{figure}[h!]
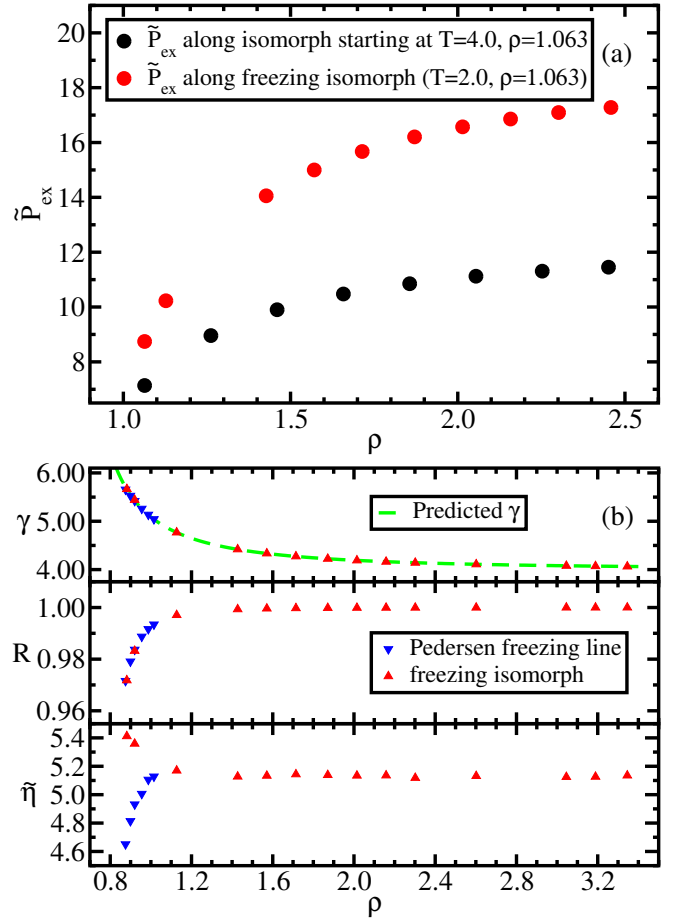

\centering
  \resizebox{9cm}{!}{
  \begin{tabular}{l}
  \begin{lpic}[t(0.5mm)]{red_pressure_liquids_vsRho(0.50)}\end{lpic} \\
  \begin{lpic}[]{freezing_red_visco_and_R_ver2_rho(0.50)}\end{lpic}
  \end{tabular}}
  \captionof{figure}{
  (a) Excess pressure in reduced units, $\tilde{P}_{ex} = W/(N k_B T)$ along two different isomorphs, 
  the freezing isomorph and a liquid isomorph. 
  For inverse power-law pair potentials this quantity is invariant, while for the LJ system it clearly is not.
  This shows that isomorph scaling is not simply a trivial IPL scaling.
  (b) In the top panel, the scaling coefficient $\gamma$, Eq. (\ref{eq:gamma}), 
  is shown as a function of density along the freezing line and the freezing isomorph. 
  The green line is the predicted value from $\gamma = d \ln h(\rho) / d \ln \rho$\cite{hrhoLasse, hrhoTrond}. 
  The middle and bottom panels show the virial potential-energy correlation coefficient $R$ and the reduced viscosity 
  $\tilde{\eta}$ along the freezing line and the freezing isomorph.
  The blue symbols mark data at freezing state points taken from Pedersen\cite{ulfpinning}; 
  the red symbols are the same quantities calculated at freezing isomorph state points. }
  \label{fig:redPandVisco}
\end{figure}

\subsection{Viscosity along the freezing line and the Andrade equation} \label{subsec:visco}
In order to evaluate the viscosity the system was simulated using the SLLOD algorithm\cite{EvansHeyes1990} 
(details are given in the Appendix). 
Studies of the viscosity of the LJ system were done in the past, e.g., by Ashurst and Hoover \cite{Ashurst1975} and 
more recently by Galliero \emph{et al}\cite{Galliero2011} and Delage-Santacreu \emph{et al}\cite{Delage2015}, 
in all cases for densities fairly close to unity.

The isomorph theory predicts the reduced viscosity to be constant to a good approximation along an isomorph 
(and therefore along the freezing line),
\begin{equation} \label{eq:redvisco}
\begin{aligned}
\tilde{\eta}\equiv\frac{\eta}{\rho^{2/3} \sqrt{m k_B T}}=\text{Const.}
\end{aligned}
\end{equation}
From this equation it is clear that if we know the value of $\eta$ at a given state point 
we can calculate the expected viscosity at any state point on the same isomorph.
Along the freezing line (F) this equation can be written as
\begin{equation} \label{eq:Freezingvisco}
\begin{aligned}
\eta_F(\rho) = \tilde{\eta}_0\cdot\rho^{2/3}\sqrt{m k_B T_F(\rho)}& 
\end{aligned}
\end{equation}
where the subscript $F$ stands for freezing, $T_F(\rho)$ is the freezing temperature at density $\rho$ 
and $\tilde{\eta}_0 = 5.2$ is the reduced value of $\eta$ at the reference state point $(\rho_0, T_0)=(1.063, 2.0)$.
Equation (\ref{eq:Freezingvisco}) is identical to the Andrade equation for the freezing viscosity 
\cite{Andrade1934a, Andrade1934b} from 1934:
\begin{equation} \label{eq:andrade}
\begin{aligned}
\eta(\rho_F, T_F)=\beta \cdot \rho_F^{2/3} \sqrt{T_F}
\end{aligned}
\end{equation}
where $\rho_F$ is the density at freezing. 
This is well known to apply for most metals to a good approximation\cite{Kaptay2005}.
The parameter $\beta$ in Eq. (\ref{eq:andrade}) depends on the system, just as the value of $\tilde{\eta}_0$ 
in Eq. (\ref{eq:Freezingvisco}) depends on the chosen potential.

\begin{figure}
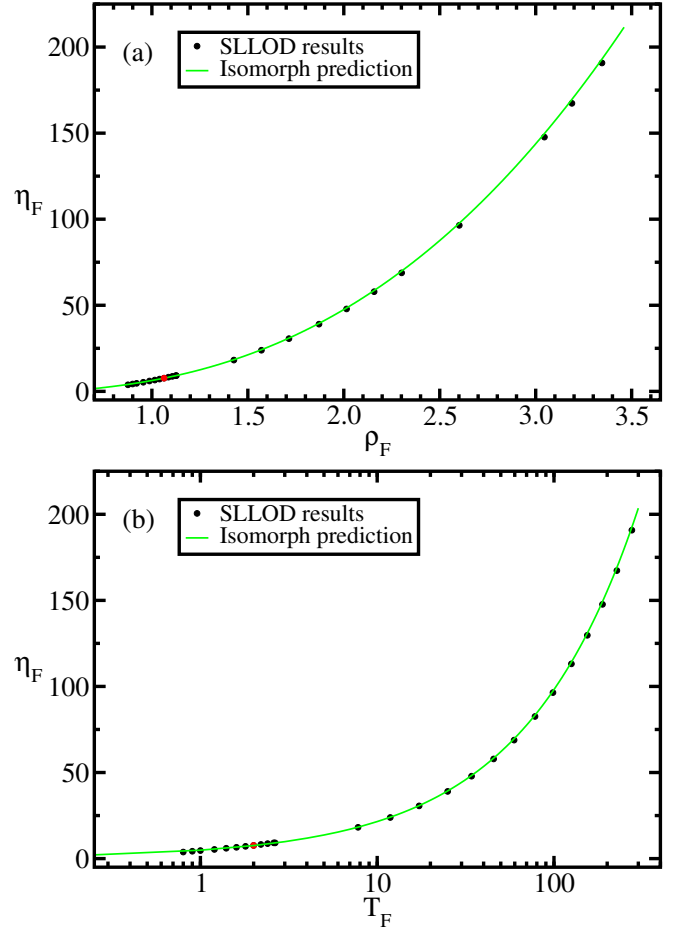

\centering
  \resizebox{9cm}{!}{
  \begin{tabular}{l}
  \begin{lpic}[]{ViscosityComplete_rho(0.50)}\end{lpic} \\
  \begin{lpic}[]{ViscosityComplete(0.50)}\end{lpic}
  \end{tabular}}
  \caption{Viscosity along the approximate freezing isomorph, Eq. \ref{eq:T(rho)2}, as a function of density (a) and 
  of temperature (b). The black dots represent results for the viscosity obtained from our SLLOD simulations (Appendix). 
  The green line is the predicted viscosity assuming the invariance of reduced viscosity along an isomorph 
  (Eq. (\ref{eq:Freezingvisco})). 
  The red dot is the viscosity of the state point from which the freezing isomorph is built and 
  the constant of Eq. (\ref{eq:Freezingvisco}) determined, $(\rho, T)=(1.063, 2.0)$. 
  The reduced viscosity at this state point is $\tilde{\eta}_0 = 5.2$}
  \label{fig:viscosityFreezing}
\end{figure}

In Fig. \ref{fig:viscosityFreezing} viscosity results are compared to the values of the viscosity predicted from 
isomorph theory using Eq. (\ref{eq:Freezingvisco}).

The green line in Fig. \ref{fig:viscosityFreezing} (b) is obtained by solving Eq. (\ref{eq:T(rho)2}) with respect to $\rho^2$ 
and using the solution to remove the $\rho$ dependence from Eq. (\ref{eq:Freezingvisco}). This results in
\begin{equation} \label{eq:visco_explicit}
\begin{aligned}
\eta(T_F)=\tilde{\eta}_0 \sqrt{m k_B T_F} \left( \frac{B_F + \sqrt{B_F^2 + 4 A_F \cdot T_F}}{2 A_F} \right)^{1/3}
\end{aligned}
\end{equation}
in which $A_F=2.27$ and $B_F=0.80$ are the freezing isomorph coefficients identified in Sec. \ref{sec:melting} 
using Eqs. (\ref{eq:h(rho)2}) and (\ref{eq:Rgammavalues}), i.e., 
based exclusively on simulations at the reference state point $(\rho, T)=(1.063, 2.0)$ 
(the units of $A_F$ and $B_F$ are $\sigma^{12} \cdot \epsilon/k_B$ and $\sigma^6 \cdot \epsilon/k_B$, 
with $\epsilon$ and $\sigma$ being the LJ parameters).
The red dot in Fig. \ref{fig:viscosityFreezing} marks the reference state point.
\begin{figure}
\centering
  \includegraphics[width=8cm]{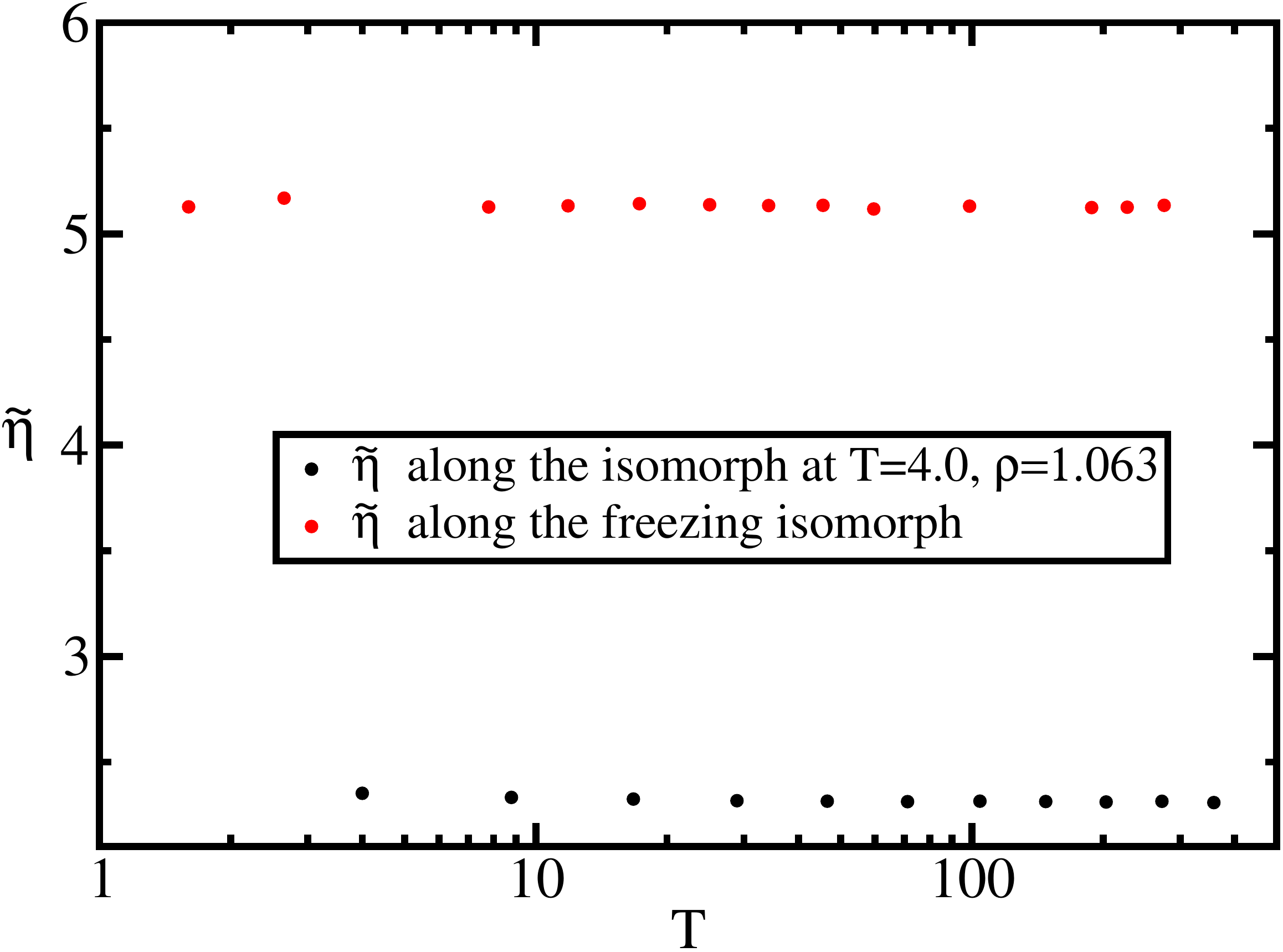}
  \caption{Reduced viscosity along the freezing isomorph and along an isomorph well within the liquid state.}
  \label{fig:redviscoiso}
\end{figure}

In Fig. \ref{fig:redviscoiso} we show the reduced viscosity along the freezing isomorph as well as along the liquid isomorph 
with reference state point $(\rho_0,T_0)=(1.063,4.0)$.
The figure demonstrates that invariance of the reduced viscosity along the freezing line is not a specific property of the freezing line, 
but a consequence of the more general isomorph invariance.

Andrade's equation for the freezing viscosity, which is explained by the isomorph theory,
was also discussed recently by Fragiadakis and Roland \cite{Fragiadakis2011}.
It is interesting to compare the temperature range accessible to experiments with that of the present work.
Fragiadakis and Roland\cite{Fragiadakis2011} reported data on liquid Argon in a range of temperatures 
corresponding to $\left[0.75, 4.17\right]$ in LJ units. 
This is impressive, but simulations allow one to cover an even wider range of freezing temperatures.

\section{Invariants along the melting line} \label{sec:resultsmelting}
\begin{figure*}
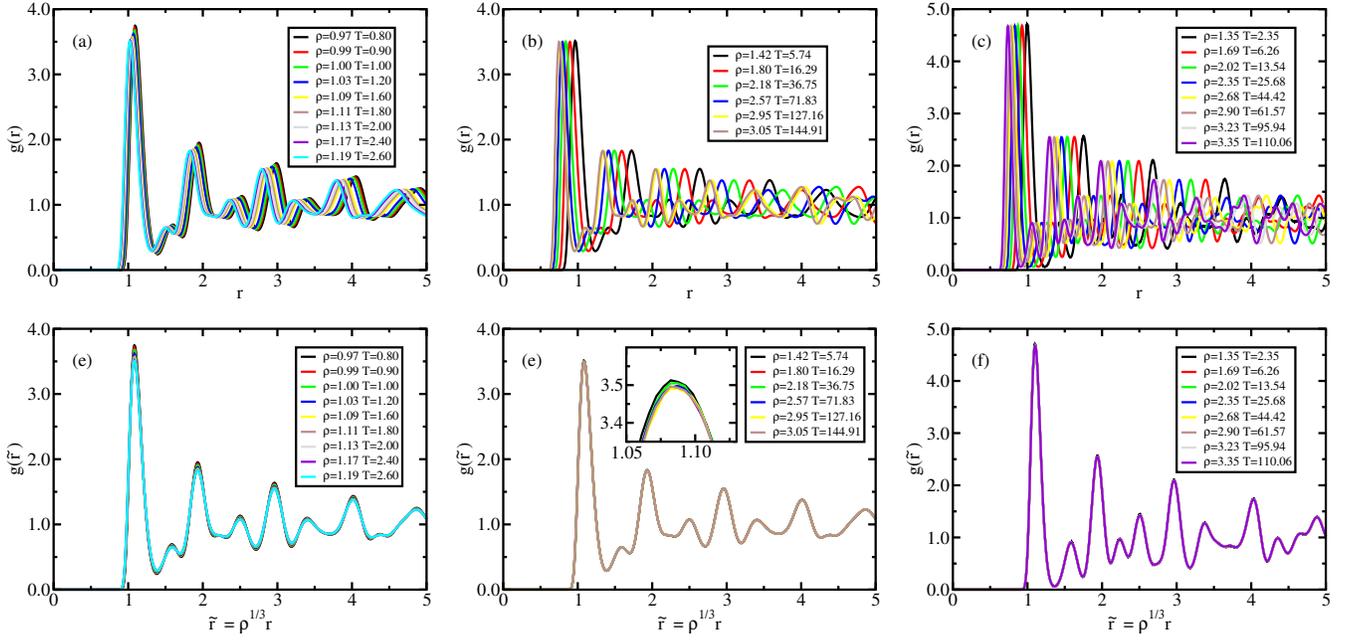

\centering
  \resizebox{18cm}{!}{
    \begin{tabular}{l l l}
  \begin{lpic}[]{ulfcrystal_rdf(0.30)}\end{lpic} & 
  \begin{lpic}[]{cryMelt_rdf(0.30)}\end{lpic} &
  \begin{lpic}[]{deepCrystal_rdf(0.30)}\end{lpic} \\
  \begin{lpic}[]{ulfcrystal_rdf_reduced(0.30)}\end{lpic} &
  \begin{lpic}[]{cryMelt_rdf_reduced(0.30)}\end{lpic} &
  \begin{lpic}[]{deepCrystal_rdf_reduced(0.30)}\end{lpic} \\
    \end{tabular}}
  \caption{Crystal results. Radial distribution function along the Pedersen melting line (a,d)\cite{ulfpinning}, 
  along the approximating melting isomorph (b,e), and along an isomorph well within the crystalline state (c,f); 
  in (a), (b), and (c), the RDFs are plotted as a function of distance in Lennard-Jones units,
  in (d), (e), and (f), the RDFs are plotted as a function of reduced distance.
  It is worth noting that while in (a) and (d) the density change is only a few percent, 
  in the other figures density is changed of about a factor $3$. The same holds for Figs. \ref{fig:MSDMelting}.}
  \label{fig:RDFMelting}
\end{figure*}

\begin{figure*}
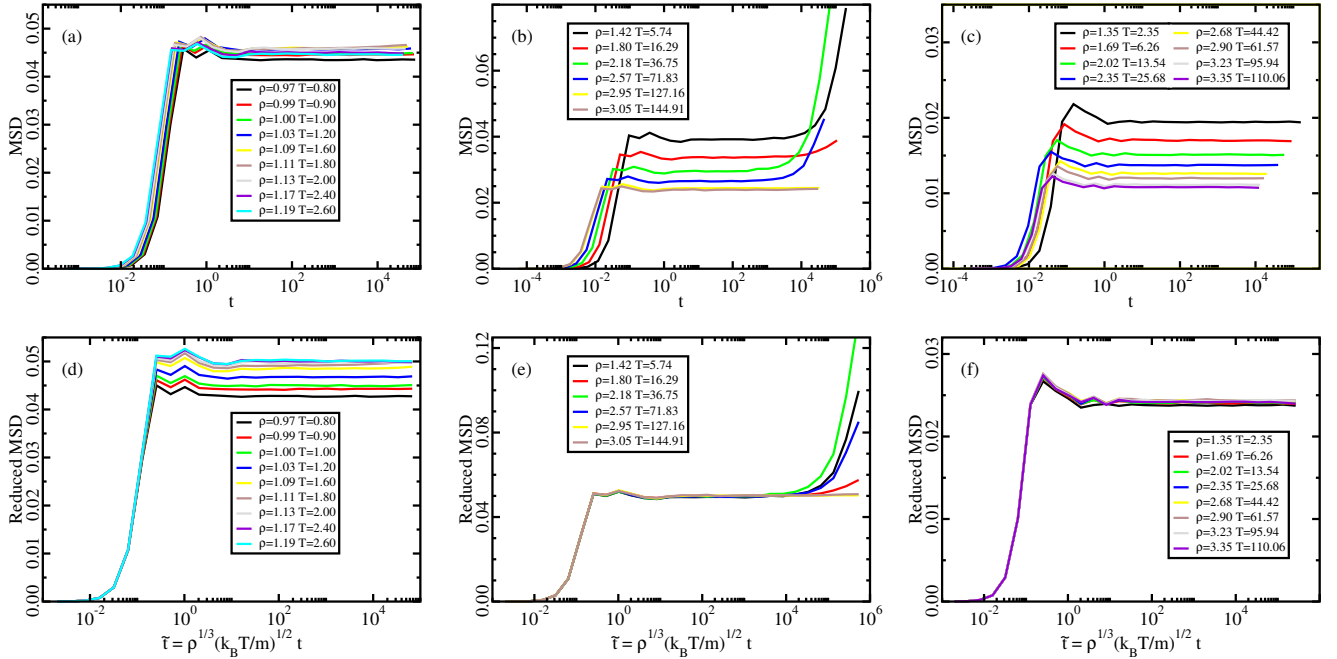

\centering
  \resizebox{18cm}{!}{
    \begin{tabular}{l l l}
  \begin{lpic}[]{ulfcrystal_msd(0.30)}\end{lpic} &
  \begin{lpic}[]{cryMelt_msd(0.30)}\end{lpic} &
  \begin{lpic}[]{deepCrystal_msd(0.30)}\end{lpic} \\
  \begin{lpic}[]{ulfcrystal_msd_reduced(0.30)}\end{lpic} &
  \begin{lpic}[]{cryMelt_msd_reduced(0.30)}\end{lpic} &
  \begin{lpic}[]{deepCrystal_msd_reduced(0.30)}\end{lpic} \\
   \end{tabular}}
  \caption{Crystal results. Mean-squared displacement along the Pedersen melting line (a,d)\cite{ulfpinning}, 
  along the approximate melting isomorph (b,e), and along an isomorph well within the crystalline state (c, f);
  in (a), (b), (c), the MSDs are plotted as a function of time in LJ units,
  in (d), (e), (f), the reduced MSDs are plotted as a function of reduced time.
  The invariance of the plateau of MSD along the melting line implies the Lindemann melting criterion for R liquids because 
  the invariance of the reduced-unit vibrational mean-square displacement in equivalent to the invariance of Lindemann constant 
  (Sec. \ref{sec:resultsmelting}).
  Along the melting isomorph defects' diffusion is observed.
  Defect formation is a stochastic phenomenon, 
  as shown by the not-monotonicity of its appearance with respect to $T$ or $\rho$.
  In order to study the isomorphic invariance of defects' formation, 
  it's necessary to average over many simulations at every state point and it could be object of future studies.
  The diffusion of defect in crystal, when properly averaged, have been shown to be isomorphic invariant 
  by Albrechtsen \emph{et al} \cite{IsomorphCrystal}.}
  \label{fig:MSDMelting}
\end{figure*}

Following the same argument as for the freezing line (Sec. \ref{sec:melting}), 
the melting line is also an approximate isomorph.
A study similar to that of Sec. \ref{sec:resultsfreezing} was performed, evaluating the structure and MSD, 
for an FCC LJ crystal along the melting line as well as another isomorph in the crystalline phase.
The starting point for the melting isomorph is taken from Pedersen \cite{ulfpinning}; 
this is the state point $(\rho, T)=(1.132, 2.0)$.
The starting point for the crystal isomorph is $(\rho, T)=(1.132, 1.0)$, which is well within the crystalline phase. 
The melting isomorph equation for the LJ system is
\begin{equation} \label{eq:meltingiso} 
T_M(\rho)=A_M\rho^4-B_M\rho^2
\end{equation}
where $A_M=1.76$ and $B_M=0.69$. 
The equation has the same mathematical form as the freezing equation, Eq. \ref{eq:T(rho)2}, (but different coefficients) 
because the shape of isomorphs reflects the pair potential, not the phase.
The existence of isomorphs in the crystalline phase was demonstrated 
in a recent publication by Albrechtsen \emph{et al} \cite{IsomorphCrystal}; 
this paper showed that the isomorph theory, in fact, is more accurate in the crystalline phase than for liquids.
In Table \ref{tab:melting} the predicted melting temperature at density $3.509$ from Eq. (\ref{eq:meltingiso}) is compared to the results for the melting line obtained in the present work using Pedersen's interface pinning method\cite{ulfpinning,pinningmethod}.
\begin{table}[h]
\small
  \caption{Comparison between the melting temperature at a given density, predicted using Eq. (\ref{eq:meltingiso}), 
  and that calculated for the same density using interface pinning method\cite{pinningmethod}. 
  The freezing and melting state temperatures at $\rho=3.509$ have been calculated in this work while the other data
  are from Pedersen\cite{ulfpinning}. 
  The parameters in Eq. (\ref{eq:meltingiso}) were calculated at the reference state point $(\rho, T)=(1.132, 2.0)$}
  \label{tab:melting}
  \resizebox{9cm}{!}{
  \begin{tabular*}{0.5\textwidth}{@{\extracolsep{\fill}}llll}
    \hline $\rho_{M}$ & $T_{M}$ & $T_{pinning}$ & $\Delta T / T_{M}$ \\
    \hline $ 0.973$	& $ 0.800$ & $ 0.921$ & $-0.132$ \\
    \hline $ 0.989$	& $ 0.900$ & $ 1.006$ & $-0.106$ \\
    \hline $ 1.005$	& $ 1.000$ & $ 1.095$ & $-0.086$ \\ 
    \hline $ 1.034$	& $ 1.200$ & $ 1.270$ & $-0.055$ \\ 
    \hline $ 1.061$	& $ 1.400$ & $ 1.453$ & $-0.036$ \\ 
    \hline $ 1.087$	& $ 1.600$ & $ 1.636$ & $-0.022$ \\
    \hline $ 1.109$	& $ 1.800$ & $ 1.812$ & $-0.007$ \\ 
    \hline $ 1.132$	& $ 2.000$ & $ 2.000$ & $+0.000$ \\
    \hline $ 1.153$	& $ 2.200$ & $ 2.191$ & $+0.004$ \\
    \hline $ 1.172$	& $ 2.400$ & $ 2.371$ & $+0.012$ \\
    \hline $ 1.191$	& $ 2.600$ & $ 2.561$ & $+0.015$ \\
    \hline $ 3.509$ & $258.44$ & $275.81$ & $+0.067$ \\
    \hline 
  \end{tabular*}}
\end{table}
As for the liquid-state isomorphs and the freezing line, the RDF is invariant both along the melting isomorph and along the crystal isomorph when expressed as a function of the reduced pair distance (Fig. \ref{fig:RDFMelting}).
The MSD is shown in Fig. \ref{fig:MSDMelting}.
The plateau of the MSD at melting confirms pressure invariance of the Lindemann melting criterion\cite{Lindemann1910,Gilvarry1956,Ross1969}.
The approximate invariance of reduced-unit MSD in the crystal implies that the value of the plateau 
for the mean atomic displacement is constant in reduced units along an isomorph (and consequently along the melting line), 
is consistent with the Lindemann criterion.
At low densities the invariance of the MSD plateau is violated.
This is the region where the melting isomorph provides a worse approximation to the LJ melting line, Fig. \ref{fig:MSDMelting}(d),
as also shown by Pedersen\cite{ulfpinning}. 
The Lindemann constant increases slightly with increasing density along melting, as reported by Luo \emph{et al}\cite{Luo2005}.
For temperatures above $1.8$, the Lindemann criterion is accurately satisfied, i.e., 
the reduced vibrational mean-square displacement becomes density independent, Figs. \ref{fig:MSDMelting} (d) and \ref{fig:MSDMelting} (e).

\section{Discussion} \label{sec:conclusions}
We have studied several properties of the LJ model along its freezing and melting lines, 
as well as along isomorphs well within the liquid and the crystalline phases.
In Table \ref{tab:allparams} the coefficients describing the four isomorphs studied in this work are given together 
with the relative reference state points.
\begin{table}[h]
\small
  \caption{This table gives the coefficient $A$ and $B$ of the isomorph equation (\ref{eq:T(rho)2}) 
  for the four isomorphs studied in this work.
  The first two columns contain the coefficients and the latter four columns contain temperatures, densities,
  density scaling coefficient $\gamma$, and correlation coefficient $R$ of the state points 
  the isomorphs studied in this work start from. A pure $n=12$ IPL pair potential leads to $\gamma=4$.}
  \label{tab:allparams}
  \resizebox{9cm}{!}{
  \begin{tabular*}{0.5\textwidth}{@{\extracolsep{\fill}}lllllll}
    \hline   & $A$ & $B$ & $T$ & $\rho$ & $\gamma$ & $R$ \\
    \hline liquid isomorph   & 4.32 & 1.34 & 4.0 & 1.063 & 4.7589 & 0.9966 \\
    \hline freezing isomorph & 2.27 & 0.80 & 2.0 & 1.063 & 4.9079 & 0.9955 \\
    \hline melting isomorph  & 1,76 & 0.69 & 2.0 & 1.132 & 4.8877 & 0.9985 \\
    \hline crystal isomorph  & 0.91 & 0.39 & 1.0 & 1.132 & 4.9979 & 0.9986 \\
    \hline
  \end{tabular*}}
\end{table}
The aim was not primarily to report that these invariances hold, which is already well known
\cite{Lawson2009,Stacey1977,Hansen1969,Ubbelohde1965,Andrade1931} albeit over smaller melting temperature / density ranges than studied here, but to relate these invariances to the isomorph theory. 
With this goal in mind we investigated whether the invariants, thought to be peculiar to the freezing/melting process, 
hold also along other isomorphs in the liquid and crystalline phase. The results show that this is indeed the case.
This means that these invariants are consequences of the LJ system being an R liquid in the relevant part of its phase diagram,
not a specific property of freezing or melting.
Nevertheless it should be stressed that invariances of reduced units quantities, 
which would be exact if the freezing/melting lines were perfect isomorphs, are violated somewhat close to the triple point.

Before discussing our results in detail, we want to point out the differences between isomorph theory and other approaches 
often used to describe the LJ system invariances.
These other attempts to describe LJ invariances are the well know hard sphere (HS) paradigm and 
the WCA (Weeks, Chandler, Andersen) approximation.
HS and isomorph theory are able to describe the nature of the same invariances, but with some important differences.
A first difference is in the possibility of determine when the theory is expected to work and when not. 
In the case of isomorph theory there is a simple prescription: if the system is strongly correlating then it is possible to build 
isomorphs along which many reduced quantities are invariant.
In the framework of hard spheres it is not possible to proceed in this way. 
It is not even possible to know from one single state point if some invariances will hold in the region around that state point 
because there is no equivalent of the correlation coefficient $R$ defined in Eq. (\ref{eq:Req}).
Another fundamental difference between the two approaches is the presence of an \textit{ad hoc} defined hard sphere radius 
that is in general state-point dependent.
Isomorph theory works without need of introducing any \textit{ad hoc} parameters. 
A last difference, which is perhaps the most important, lies in the possibility of predicting which invariances the system will have.
According to the HS paradigm, once the mapping from the studied system to the HS system is done using the 
\textit{ad hoc} defined HS radius, the invariances of the HS system are inherited from the studied system. 
This means that structure, dynamics and thermodynamic quantities should be invariant along constant-packing-fraction curves. 
In Fig. \ref{fig:redPandVisco} we showed that the reduced pressure of the LJ system is not invariant along an isomorph (a) 
while reduced viscosity is (b), as predicted from isomorph theory.
Another possible comparison is between the isomorph theory and the WCA approximation for the LJ system.
While in isomorph theory there is no reference system, 
the WCA approximation is based on the idea that only the repulsive part of the LJ potential is relevant in the description 
of the system, providing a convenient reference system, and that LJ invariances can be derived from HS invariances\cite{Andersen1971}.
%\begin{figure}
%\centering
%  \includegraphics[width=8cm]{plot_fullvisco_ver2.pdf}
%  \caption{
%  Viscosities (inset) of IPL12 system and of LJ system along the freezing line (data from Agrawal and Kofke\cite{Agrawal1995}) 
%  and their ratio (main figure).
%  The viscosities are calculated using the SLLOD algorithm\cite{EvansMorriss2008, DaivisTodd2006, EvansHeyes1990}.
%  The viscosity of the IPL12 system is substantially different from that of the LJ system for temperatures lower than 
%  $T=68.5$ in LJ units.}
%  \label{fig:IPLvsLJ}
%\end{figure}
\begin{figure}[t!]
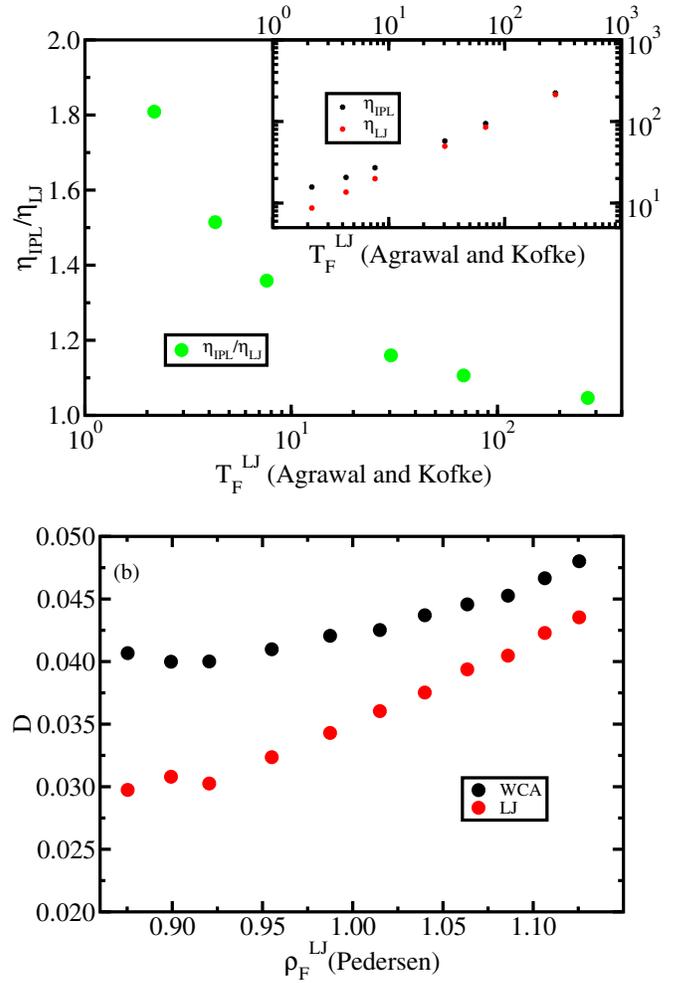

\centering
  \resizebox{9cm}{!}{
  \begin{tabular}{l}
  \begin{lpic}[]{plot_fullvisco_ver2(0.50)}\end{lpic} \\
  \begin{lpic}[]{WCA_LJ_diffusion(0.50)}\end{lpic}
  \end{tabular}}
  \caption{
  (a) Viscosities (inset) of IPL12 system and of LJ system along the freezing line 
  (data from Agrawal and Kofke\cite{Agrawal1995}) and their ratio (main figure).
  The viscosities are calculated using the SLLOD algorithm\cite{EvansMorriss2008, DaivisTodd2006, EvansHeyes1990}.
  The viscosity of the IPL12 system is substantially different from that of the LJ system for temperatures lower than 
  $T=68.5$ in LJ units.
  (b) Diffusion constant for the LJ system and WCA system along the Pedersen freezing line. 
  It is well known that the WCA potential reproduces with good accuracy the structure of the LJ system while 
  this is not the case for dynamics, as the figure shows.}
  \label{fig:IPLvsLJ}
\end{figure}

In Fig. \ref{fig:IPLvsLJ} (a) the viscosity is shown along the freezing line data from Agrawal and Kofke\cite{Agrawal1995} 
for the LJ system and for the IPL potential:
\begin{equation} \label{eq:visco_explicit}
\begin{aligned}
v_{IPL}(r)=4 r^{-12}
\end{aligned}
\end{equation}
which is the repulsive term of the LJ potential. 
The viscosity calculated with the two different potentials along the freezing line is very different.
The difference is larger than $10\%$ before reaching considerably high densities and temperatures 
($(\rho, T)=(2.417, 68.5)$ in LJ units). 
This means that the effects of attraction are not negligible up to really high densities.
As Rosenfeld wrote in $1976$ \emph{"It is important here to emphasize that the $r^{-6}$ term of the L-J potential gives 
appreciable contribution to the thermodynamic properties of the system up to very high temperatures"}\cite{Rosenfeld1976b} 
regarding the difference between the freezing line of IPL12 and LJ.

In Fig. \ref{fig:IPLvsLJ} (b) the diffusion constant $D$ for the LJ system with WCA approximation and with the 2.5 cutoff are shown. The WCA approximation is well known to reproduce with good accuracy the structure of the LJ system, 
but it fails in reproducing the dynamics. 
Berthier and Tarjus\cite{Berthier2009} already underlined that this was the case for Kob-Andersen binary LJ system and 
Pedersen \emph{et al}\cite{ulf2010} showed how isomorph theory provide a better description of the LJ system dynamics 
while preserving the good description of structure.

In Secs. \ref{sec:resultsfreezing} and \ref{sec:resultsmelting} we discussed the relation 
between the isomorph theory and freezing/melting criteria.
It was shown that the invariance along the freezing line of the maximum of the static structure factor $S(q)$ 
(the Hansen-Verlet criterion) results from a general invariance along isomorphs of the entire $S(q)$ function. 
The first peak of the structure factor along an isotherm decreases gradually with decreasing density. 
This means that there will be a specific value which corresponds to the freezing phase transition.
The evidence that the value of this height is constant along the freezing line is not a peculiarity of the
freezing process itself, but a consequence of isomorph scaling. 
The reason why the maximum height of $S(q)$ is $2.85$\cite{Hansen1969,Hansen1970} cannot be explained within isomorph theory, 
but is a feature of the freezing process.
In order to explain the universality of the number $2.85$, as well as the universality of the Lindemann melting criterion number, 
one must refer to quasiuniversality, a further consequence of the isomorph theory detailed, e.g., 
by Bacher \emph{et al}\cite{Bacher2014NatCom}.
Note the compatibility of the general isomorph theory with the results of Saija \emph{et al}\cite{Saija2006} 
on the pair-potential dependence of the maximum height of $S(q)$ at freezing.

The study of the LJ structure factor along the freezing line allows also to explain some 
properties of structure factors for liquid metals observed in X-rays experiments. 
As shown by Waseda and Sukuri in 1972 \cite{Waseda1972}, for some liquid metals the ratio of the position of the first and second peak 
in the structure factor is the same and others for which this does not hold, as for example Ga, Sn, Bi. 
The first set of metallic liquids are the ones which are R liquids (i.e., exhibit strong virial potential-energy correlations), 
and therefore are similar to the LJ system studied in this work, while those in the second do not, 
as shown very recently by Hummel \emph{et al} \cite{Hummel2015} from \emph{ab initio} density functional theory calculations.

Along the melting line we studied the Lindemann criterion, which has been widely discussed
\cite{Saija2006,Luo2005,Lawson2009,Lowen1993} and also experimentally tested\cite{Sokolowski2003}, 
and the same conclusion holds as for the Hansen-Verlet criterion.
Isomorphs' existence implies that an R liquid's thermodynamic phase diagram becomes effectively one-dimensional 
with respect to the isomorph-invariant quantities. 
The reduction of the 2d phase diagram to an effectively 1d phase diagram is crucial for understanding the connection between 
the isomorph theory and the Lindemann criterion, because it removes one of the main criticism against this criterion, 
i.e., it being a single-phase criterion \cite{Ubbelohde1965}. 
If the phase diagram is effectively one-dimensional, there is a unique melting process and the Lindemann constant 
is the value associated with this phase transition; the invariance of Lindemann constant along the melting line is, 
in this view, a consequence of isomorph invariance.
This argument also explains why one can use a single-phase criterion to predict where the 
melting process takes place for R liquids.
According to the Lindemann criterion, the crystal melts when the vibrational MSD exceeds a threshold value, 
which in reduced units is constant along the melting line. 
This condition is equivalent to the invariance of the MSD along the melting line, an isomorph prediction.
Note that the isomorph theory can be used to predict for which systems the Lindemann criterion (at least) must hold, namely all R liquids.
Recent comprehensive density-functional theory (DFT) simulation data of Hummel \emph{et al}\cite{Hummel2015} show that 
most metals are R liquids and therefore the Lindemann criterion must apply for them in the sense 
that the reduced-unit MSD is approximately invariant along the melting line. 
On the other hand, systems that do not exhibit strong correlations between virial and potential-energy do not necessarily obey 
the Lindemann criterion. 
Thus as discussed by Stacey and Irvine already in 1977 \cite{Stacey1977}, 
the Lindemann criterion applies for systems which ``undergo no dramatic changes in coordination on melting''. 
This is not the case for hydrogen-bonding systems, which are not R liquids\cite{paper1,simpleliquid}.
The non-universal validity of the Lindemann criterion is also supported by Lawson\cite{Lawson2009} and 
by Fragiadakis and Roland\cite{Fragiadakis2011}. Another interesting point is the connection between the Lindemann and Born criteria,
relating melting to the vanishing of the shear modulus in the crystal.
Jin \emph{et al}\cite{Jin2001} showed that for a LJ system when the Lindemann criterion is satisfied, 
the Born criterion\cite{Born1939} holds too to a good approximation. 
In view of the isomorph theory this is not surprising, because the reduced shear modulus is invariant along an isomorph 
and therefore constant on melting.

In Sec. \ref{sec:resultsfreezing} we discussed the relation between the isomorph theory and Andrade's viscosity equation 
from 1934 for the viscosity of liquid metals at freezing. 
This equation is equivalent to stating invariance of the reduced viscosity along an isomorph, 
Eqs. (\ref{eq:Freezingvisco}) and (\ref{eq:andrade}).
As for the Lindemann criterion, the isomorph theory provides the possibility to predict whether a liquid will obey Andrade equation. 
The Hummel \emph{et al} \cite{Hummel2015} DFT simulation data explain why this equation holds 
for liquid alkali metals (as well as other invariances\cite{Balucani1993}); 
likewise one also expects this equation to hold for many other metals, for example iron. 
This last point is of significant interest because the estimation of viscosity of liquid iron close to freezing 
in the Earth core is of crucial relevance for the development of Earth-core models 
\cite{Poirier1988, Poirier2000,Stacey2010}, but still widely debated\cite{Rutter2002,Fomin2013,Shen2004}. 
Isomorph scaling predicts an increase of the real (non-reduced) viscosity along the freezing line
consistent with the results of Fomin \emph{et al}\cite{Fomin2013}.
\begin{figure}[t!]
\centering
  \includegraphics[width=8cm]{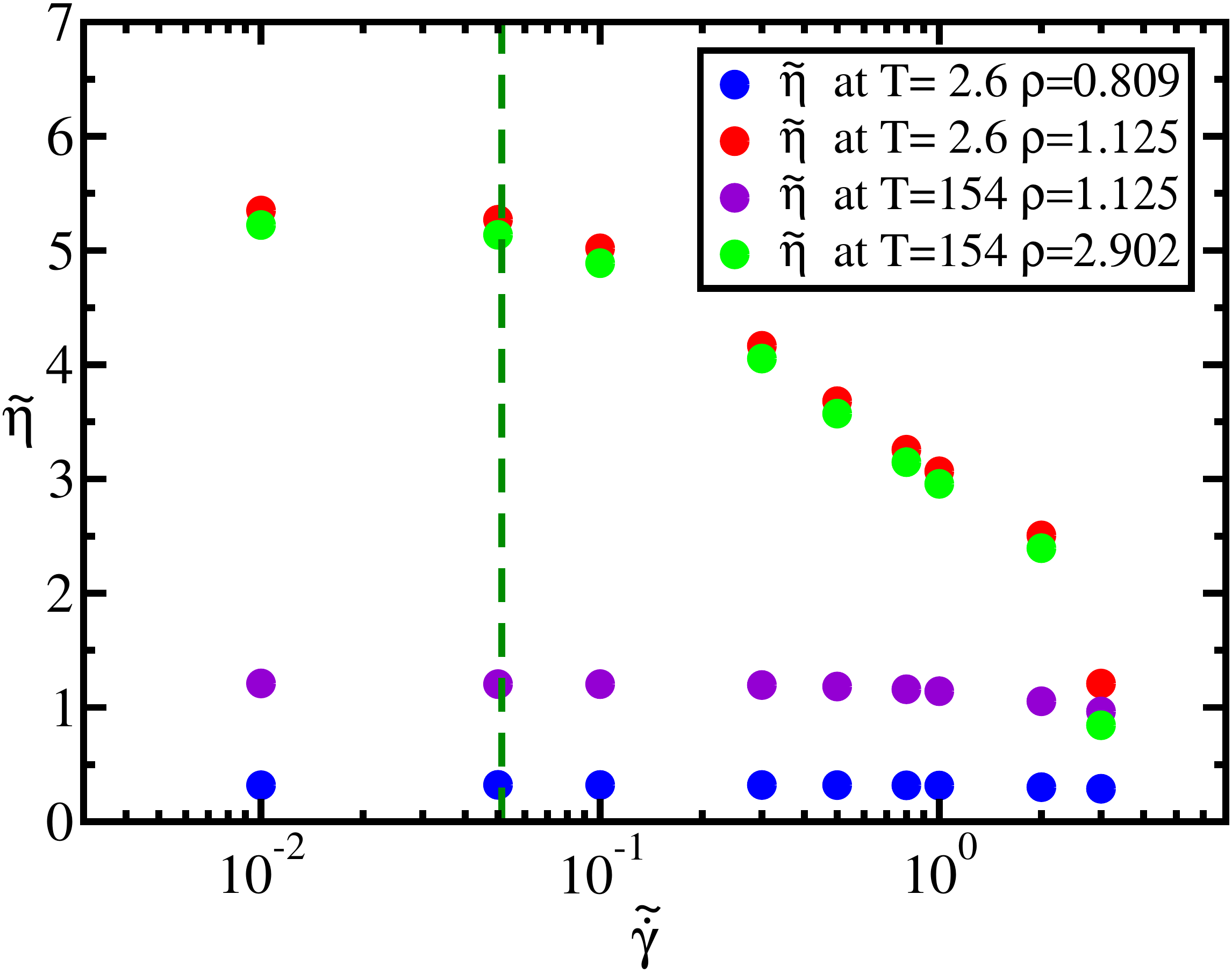}
  \caption{Measured reduced viscosity from equation \ref{eq:SLLODvisco} at different reduced strain rates.
  Two of the five state points (green and red dots) are on the same isomorph (the freezing isomorph of Sec.
  \ref{sec:melting}) and their behavior in reduced units is the same. 
  As consequence, the reduced strain rate at which reduced viscosity start to be strain rate dependent 
  is isomorphic invariant, consistently with results from Separdar\emph{et al}\cite{sllod}.
  The blue and violet dots are results of simulations at state points isochoric or isothermic to the $(\rho,T)=(1.125,2.6)$. 
  The behavior of reduced viscosity as function of strain rate is strongly modified changing density
  or temperature if the chosen state points are not isomorphic.}
  \label{fig:viscolin}
\end{figure}

\section{Conclusions}
We have shown that the freezing and melting lines are approximately isomorphs, 
how the isomorph theory can be used to explain why some liquids have simple behavior at freezing and melting, 
i.e., have several structural and dynamical approximate invariants along the freezing and melting lines. 
Thus this theory can be used for R liquids to determine melting and freezing physical quantities not easily accessible by experiments, 
ranging from noble gasses like Argon over liquid metals to certain molecular liquids. 

\section*{Acknowledgements}
The authors are grateful to Ulf R. Pedersen and Nicholas P. Bailey for many useful discussions.
The center for viscous liquid dynamics \emph{Glass and Time}
is sponsored by the Danish National Research Foundation via Grant No. DNRF61. 

\appendix
\section{Determining the zero-strain rate viscosity from SLLOD simulations}

A SLLOD simulation\cite{EvansMorriss2008, DaivisTodd2006, EvansHeyes1990} is a molecular dynamics simulation performed 
by shearing the simulation box with constant speed.
Between the bottom part of the box and the top part there is a relative shearing motion with strain rate 
$\dot{\gamma}=\frac{\partial u_x}{\partial y}$, where $u_x$ is the streaming velocity at ordinate $y$ 
when the box is sheared in the $x$ direction.
Under low strain-rate conditions, this kind of simulation reproduces an ordinary, linear Coulette flow and the linear, 
shear-rate-independent, viscosity can be calculated from the stress tensor $\sigma_{ij}$ through the equation
\begin{equation} \label{eq:SLLODvisco}
\begin{aligned}
\eta=\frac{\sigma_{xy}}{\dot{\gamma}} 
\end{aligned}
\end{equation}
Equation (\ref{eq:SLLODvisco}) holds only when the viscosity is independent of strain rate, i.e., at a sufficiently small shear rate. 
As shown by Separdar \emph{et al}\cite{sllod} the strain rate $\dot{\gamma}$ for which the measured viscosity starts to be strain-rate dependent is isomorph invariant when given in reduced units. 

The behavior of the reduced viscosity $\tilde{\eta}$ as a function of the reduced strain rate $\tilde{\dot{\gamma}}$ 
is shown in Fig. \ref{fig:viscolin}. 
When the two considered state points are on the same isomorph, they exhibit the same shear-thinning behavior in reduced units;
this is not true if we move along an isochore or along an isotherm.
The dotted green line in Fig. \ref{fig:viscolin} marks the reduced strain rate used for the simulations along the freezing line reported in the paper.
%%%END OF MAIN TEXT%%%

%The \balance command can be used to balance the columns on the final page if desired. It should be placed anywhere within the first column of the last page.

%\balance

%If notes are included in your references you can change the title from 'References' to 'Notes and references' using the following command:
%\renewcommand\refname{Notes and references}

%%%REFERENCES%%%
%% This BibTeX bibliography file was created using BibDesk.
%% http://bibdesk.sourceforge.net/

%% Created for Lorenzo Costigliola at 2015-10-19 20:11:04 +0200 

%% Saved with string encoding Unicode (UTF-8) 

%\bibliography{MyBibliografy} %You need to replace "rsc" on this line with the name of your .bib file
%\bibliographystyle{rsc} %the RSC's .bst file

\providecommand*{\mcitethebibliography}{\thebibliography}
\csname @ifundefined\endcsname{endmcitethebibliography}
{\let\endmcitethebibliography\endthebibliography}{}

\end{document}